\documentclass[modern]{rnaastex}
\pdfoutput=1

\usepackage{graphicx}
\usepackage{microtype}
\usepackage{url}
\usepackage{amsmath}
\usepackage{amssymb}
\usepackage{natbib}
\usepackage{multirow}
\usepackage{graphicx}
\usepackage{url}
\usepackage{xspace}

\newcommand{\ie}{{\textit{i.e.},~}}
\newcommand{\eg}{{\textit{e.g.},~}}
\newcommand{\equref}[1]{{\xspace}Eq.~(\ref{#1})}
\newcommand{\figref}[1]{{\xspace}Fig.~\ref{#1}}

\newcommand{\secref}[1]{{\xspace}Sec.~\ref{#1}}
\renewcommand{\d}{{\mathrm{d}}}
\newcommand{\equ}[1]{\begin{equation}#1\end{equation}}
\newcommand{\eqn}[1]{\begin{eqnarray}#1\end{eqnarray}}
\renewcommand{\vec}[1]{\boldsymbol{#1}}

\newcommand{\ntypes}{{\mathrm{N}_\mathrm{types}}}
\newcommand{\nobj}{{\mathrm{N}_\mathrm{obj}}}
\newcommand{\ncorr}{{\mathrm{N}_\mathrm{corr}}}
\newcommand{\photoz}{\xspace{photo-$z$}\xspace}

\newcommand{\code}[1]{\xspace{{\textsc{#1}}\xspace}}
\newcommand{\bpz}{\xspace{\textsc{bpz}}\xspace}
\newcommand{\skynet}{\xspace{\textsc{skynet}}\xspace}

\begin{document}\raggedbottom\sloppy\sloppypar\frenchspacing

\vspace*{-1cm}
\title{Hierarchical modeling and statistical calibration\\ for photometric redshifts} 

\author[0000-0002-3962-9274]{Boris Leistedt}
\affil{Center for Cosmology and Particle Physics, Department of Physics, New
York University, New York, NY}
\affil{NASA Einstein Fellow}

\author[0000-0003-2866-9403]{David W. Hogg}
\affil{Center for Cosmology and Particle Physics, Department of Physics, New
York University, New York, NY}
\affil{Center for Computational Astrophysics, Flatiron Institute, New York, NY 10010, USA}
\affil{Center for Data Science, New York University, New York, NY 10011, USA}
\affil{Max-Planck-Institut f\"ur Astronomie, Heidelberg, Germany}

\author[0000-0003-2229-011X]{Risa H. Wechsler}
\affil{Kavli Institute for Particle Astrophysics and Cosmology \& Physics Department, Stanford University, Stanford, CA 94305, USA}
\affil{SLAC National Accelerator Laboratory, Menlo Park, CA 94025, USA}

\author[0000-0002-0728-0960]{Joe DeRose}
\affil{Kavli Institute for Particle Astrophysics and Cosmology \& Physics Department, Stanford University, Stanford, CA 94305, USA}
\affil{Department of Physics, Stanford University, 382 Via Pueblo Mall, Stanford, CA 94305, USA}

\begin{abstract}
The cosmological exploitation of modern photometric galaxy surveys requires both accurate (unbiased) and precise (narrow) redshift probability distributions derived from broadband photometry.
Existing methodologies do not meet those requirements.
Standard template fitting delivers interpretable models and errors, but lacks flexibility to learn inaccuracies in the observed photometry or the spectral templates.
Machine learning addresses those issues, but requires representative training data, and the resulting models and uncertainties cannot be interpreted in the context of a physical model or outside of the training data.
We present a hierarchical modeling approach simultaneously addressing the issues of flexibility, interpretability, and generalization.
It combines template fitting with flexible (machine learning-like) models to correct the spectral templates, model their redshift distributions, and recalibrate the photometric observations.
By optimizing the full posterior distribution of the model and solving for its (thousands of) parameters, one can perform a global statistical calibration of the data and the SED model.
We apply this approach to the public Dark Energy Survey Science Verification data, and show that it provides more accurate and compact redshift posterior distributions than existing methods, as well as insights into residual photometric and SED systematics. 
The model is causal, makes predictions for future data (\eg additional photometric bandpasses), and its internal parameters and components are interpretable. 
This approach does not formally require the training data to be complete or representative; in principle it can even work in regimes in which few or no spectroscopic redshifts are available.
\end{abstract}

\keywords{galaxies: distances and redshifts --- galaxies: photometry --- galaxies: statistics --- large-scale structure of universe}

\section{1. Introduction}

Modern photometric surveys such as Dark Energy Survey \citep[DES,][]{Abbott:2005bi}, the Kilo-Degree Survey \citep[KIDS,][]{deJong:2013}, and the upcoming LSST \citep{Abell:2009aa}, give us the fluxes and morphologies of hundreds of millions of galaxies\footnote{
In what follows we focus on galaxies, but it should be obvious that most of the discussion and methods of this paper are equally applicable to other extragalactic objects such as quasars.}.
By extracting the statistical properties of those galaxies and confronting them with theoretical predictions, one can test and compare cosmological models in great detail \citep[see \eg][]{Peacock:2006kj, Weinberg:2012es}.
The results of those surveys will prove essential to uncover the properties of dark matter, measure the properties of high-energy particles such as neutrinos, and test models of the early universe, gravity, and the late-time accelerated expansion.

Most physical signatures of interest (\eg the clustering of matter, or gravitational lensing) are three-dimensional and evolve with cosmic time.
Therefore, cosmological analyses require estimating the redshifts of galaxies and also grouping objects in redshift bins.
More specifically, three types of estimates are typically required: 
1) an average or point estimate of the redshift of each object, 
2) the probability distribution (referred to as ``redshift PDF'' below), and 
3) the redshift distributions of populations of galaxies (the binning process may itself use the previous point estimate). 
A new challenge in the analysis of modern surveys is that all three must reach high levels of precision and accuracy. 
\textit{Precision} is typically set by the amount of data and the noise levels, which naturally decrease as larger surveys are conducted and improved technology (\eg CCDs) is developed. 
For instance, improvements in flux noise tighten the support of the redshift PDFs, and the rise in numbers of detected galaxies decreases the uncertainty in redshift distributions.
\textit{Accuracy}, by contrast, refers to the quality of the estimates, for example the bias and scatter of the best redshift estimates.
Precise but inaccurate estimates are not of much use; in practice, any residual biases must be made smaller than the uncertainties.
Photo-$z$ errors (both statistical and systematic) are the leading term in the uncertainty budget of modern surveys, partly because standard methodologies fail to deliver accurate redshift estimates. 

Photometric redshift estimation is made challenging by the unavailability of sufficiently large and representative validation data, \ie a large set of galaxies for which redshifts are known, and uniformly sampling the survey of interest in terms of all physical and observational characteristics (depth, color, morphology, redshift, etc).
Obtaining a representative sample for modern surveys is both unrealistic and unreasonable for many reasons, including the fact that spectroscopy at the required depth is prohibitively expensive and time consuming, and that complicated (and often poorly understood) selection effects are unavoidable in both the photometric and the spectroscopic surveys.
Without large representative validation data, it is difficult to calibrate templates for the spectral energy distribution (SED) of galaxies to meet \photoz accuracy requirements. 
The present work partly aims at resolving this issue.
We now outline the advantages and limitations of the two main classes of algorithms used for estimating redshifts from photometric fluxes, template fitting and machine learning methods, which our methodology takes advantage of\footnote{Another class of methods for estimating redshift distributions for photometric data exists: ``clustering redshifts" \citep[\eg][]{Matthews:2010an}.
It exploits spatial information and the proximity of galaxies in real space (sky position and redshift).
Here we focus on exploiting photometric measurements only, and we do not discuss this class of methods. 
Any flux-based \photoz method (such as the one presented below) could be improved or calibrated by adding spatial information via clustering redshifts.
}. 
More details can be found in recent \photoz studies conducted with deep surveys  \citep[see \eg][]{Newman:2013cac, Dahlen:2013fea, Sanchez:2014zgq, Schmidt:2014ela, Bonnett:2015pww}.

Template-fitting methods \citep[\eg][]{Benitez:1998br, Brammer:2008qv, Feldmann:2006wg} are based on a library of galaxy  SEDs, used to solve for the redshift and type of a galaxy given the observed photometric fluxes and the noise estimates.
This is performed in a fully probabilistic fashion, with explicit priors over the distributions of types and redshifts of those galaxy types\footnote{The types are simply the ordered and labeled or numbered set of SEDs; we do not use continuous parameterizations here. But the types could be a discretization of a physical space of interest.}.
One obtains an interpretable posterior probability distributions for redshifts and also other galaxy properties (\eg star formation history, dust, etc), based on a well-defined set of assumptions (the SEDs, the priors, the likelihood function).
However, the rigidity of template-fitting approaches is also their main limitation.
The complexity and imperfections of observed fluxes (\eg biases or underestimated errors) cannot easily be captured in this approach without a explicit modeling and comprehensive testing with external data.
Some calibration techniques exist \citep[and are part of some of the standard SED fitting packages, \eg][]{Benitez:1998br, Brammer:2008qv, Feldmann:2006wg}, but are often performed with fairly simple corrections and separately for the different parts of the models.

Machine-learning methods \citep[\eg][]{Kind:2013eka, Collister:2003cz, Sadeh:2015lsa} address SED fitting issues by directly fitting the relationship between galaxy fluxes and redshifts with a very flexible (often nonparametric) function of the data, which depends on the algorithm under consideration (\eg neural networks, random forests, etc).
This approach is powerful for learning complicated relationships in the data (biases in the photometry or the noise estimates can be learned).
For this reason, it usually provides excellent redshift estimates, possibly the best (exhausting the available information in the photometry) assuming the model is flexible enough and trained correctly.
However, this approach only works for regions of parameter or data space (not only fluxes, redshifts, and morphology, but also their relationships to systematics like artefacts and blending in the images) with training data.
In other words, machine-learning methods excel in the interpolation regime, but cannot extrapolate outside of the training data (in machine-learning jargon, they struggle to \textit{generalize}), partly because they do not know about the underlying physics of the problem (flux measurements arise from observing a redshifted SED observed through known photometric band-passes).
This is a significant issue since, as mentioned before, training sets are typically not representative (in particular, they are shallower than the target photometric surveys).

In this work, we attempt to address those limitations and develop a hierarchical SED model that has (some of) the flexibility of machine-learning methods and the interpretability and generalization features of template fitting. 
The novelty of the work essentially lies in the greater flexibility of the model, its causal and hierarchical nature, and the joint optimization of all the parameters for obtaining photometric redshifts.
Furthermore, the ability to activate or deactivate components in the hierarchical model is a new way to gain intuition about existing photo-$z$ results as well as about the models and data under consideration. 
We apply this approach on the DES Science Verification (SV) data, and develop both a physical SED model and a photometric data model that yield statistically accurate redshifts PDFs.
Our goal is not to obtain the best photometric redshift possible for these data, but to show how the limitations of existing methodologies can be elegantly addressed with hierarchical modeling, and how this approach opens new perspectives for meeting the stringent \photoz requirements of ongoing and future surveys. 

The remainder of this paper is structured as follows: In \secref{sec:methods}, we present the hierarchical \photoz inference methodology. We illustrate its performances on the DES SV data in \secref{sec:data}. We discuss the results and conclude in \secref{sec:concl}.

\section{2. Hierarchical flux--redshift modeling}\label{sec:methods}

\begin{figure}\centering
\includegraphics[width=12cm, trim = 5cm 5cm 5cm 5cm, clip]{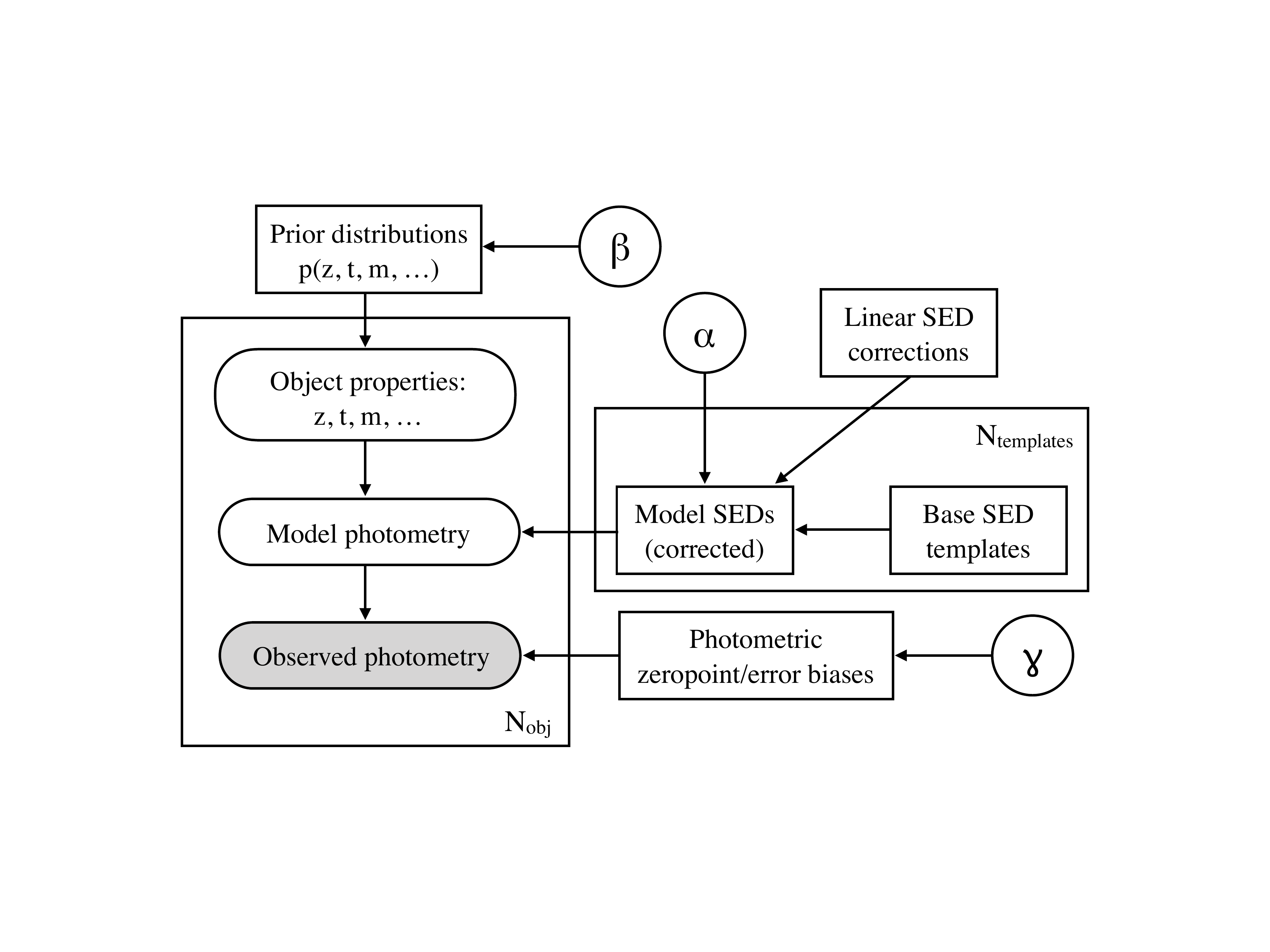}
\caption{Graphical representation of our hierarchical probabilistic model. Plates indicate independent, identically distributed variables: the observed galaxies and the model SED templates. Squares indicate fixed parameters or model components: the analytic form of the priors and the noise recalibration, as well as the linear SED bases used in the templates, their corrections and variances. Circles or round boxes refer to random variables. We optimize for the hyperparameters of the model, $\vec{\alpha}$,  $\vec{\beta}$, and $\vec{\gamma}$, and marginalize over the intrinsic true properties of the objects (true photometry, redshift, type, luminosity). The shaded rounded box refer to observed random variables, \ie the measured fluxes and their noise. Standard template fitting typically only include the left part of the model, without SED or photometric corrections, or with simple corrections not optimized simultaneously with the priors and other parameters of the model. The novelty of our approach lies in the greater flexibility of the model, its causal structure, and the joint optimization of all the parameters for obtaining photometric redshifts.}\label{fig:pgm}
\end{figure}

In this section we describe a hierarchical model for fitting noisy photometry and estimating galaxy redshifts.
The model is schematically summarized in \figref{fig:pgm} and has three main components: the spectral energy distributions, their prior distributions, and a noise recalibration module. 
Those are characterized by sets of hyperparameters denoted by $\vec{\alpha}$,  $\vec{\beta}$, and $\vec{\gamma}$, respectively.
We will apply this model to a set of objects with known redshifts and solve for the hyperparameters by maximizing the full posterior distribution.
Note that our modelling choices are relatively simple, partly because this paper is a demonstration of the approach, and these models are sufficient for the data we consider. 
In what follows, we describe the various components of the model, then write the full posterior distribution and discuss the optimization procedure.

\subsection{2.1 The likelihood}

For the $i$th galaxy, we assume that we have a vector of observed fluxes $\hat{\vec{F}}_{\hspace*{-2pt}i}$ with Gaussian noise covariance matrix $\vec{\Sigma}_{i}$, with the number of bands $B$  setting their dimensions.
The likelihood function is a multivariate Gaussian\footnote{We use the notation $\mathcal{N}(a; A)$ for a multivariate Gaussian of mean $a$ and covariance $A$.} $p(\hat{\vec{F}}_{\hspace*{-2pt}i}  | \vec{F})  =  \mathcal{N}\bigl( \vec{F} - \hat{\vec{F}}_{\hspace*{-2pt}i}; \vec{\Sigma}_{i}\bigl)$ with $\vec{F}$ the true flux vector of the object. 
We will assume that the latter is Gaussian distributed, with mean $\vec{F}^\mathrm{mod}$ and covariance $\vec{\Omega}^\mathrm{mod}$. Marginalizing over $\vec{F}$ leads to what we will operationally use as the likelihood,
\eqn{
	p(\hat{\vec{F}}_{\hspace*{-2pt}i} | \vec{F}^\mathrm{mod}, \vec{\Omega}^\mathrm{mod})  \ &=& \ \int \mathrm{d}\vec{F} p(\hat{\vec{F}}_{\hspace*{-2pt}i} | \vec{F})p(\vec{F} | \vec{F}^\mathrm{mod}, \vec{\Omega}^\mathrm{mod}) \\
	&=& \ \mathcal{N}\bigl( \vec{F}^\mathrm{mod} - \hat{\vec{F}}_{\hspace*{-2pt}i}; \vec{\Sigma}_{i} + \vec{\Omega}^\mathrm{mod} \bigl) \ =  \prod_{b=1}^B  \mathcal{N}\bigl( {F}^\mathrm{mod}_b- \hat{{F}}_{ib}; \sigma^2_{ib} + \omega^2_{b} \bigl).\quad \label{eq:datalikelihood}\nonumber
}
For the last step we made an additional assumption that the variances are uncorrelated and can be written as $\sigma^2_{ib}$ and $\omega^2_{b}$ (the diagonal terms of $\vec{\Sigma}_{i}$ and $\vec{\Omega}^\mathrm{mod}$).
Some of those vector or matrix elements could vary for each galaxy, such as the number of bands $B$, or the fact that the fluxes are uncorrelated. 
We do not consider those extensions here, but they are trivial to include.

Our mean flux model and its variance will be functions of physical parameters of interest, such as redshift, luminosity, dust, star formation rate, etc. 
In this work we will only consider galaxy type, redshift, and luminosity, and construct the model from templates of the spectral energy distributions (SEDs) of galaxies, \ie the $b$-th component of $\vec{F}^\mathrm{mod}$ will be written as ${F}^\mathrm{mod}_b(t, \ell, z)$.

\subsection{2.2 The flux--redshift model}

Our model for the photometric flux ${F}^\mathrm{mod}_b(t, \ell, z)$ for a galaxy type indexed by $t$, at redshift $z$, and luminosity $\ell$, in a band $b$, arises from integrating a template of the rest-frame flux density $f_t(\lambda)$, redshifted, multiplied with the response of the photometric filter $W_b(\lambda)$, and scaled with the luminosity $\ell$.
Thus, we have
\eqn{
	 {F}^\mathrm{mod}_b(t, \ell, z) = \ell {F}_{tb}(z) = \ell \frac{1+z}{g^\mathrm{AB}4\pi D^2_L(z)} \frac{ \int \d \lambda f_t(\lambda/(1+z)) W_b(\lambda) }{ \int \d \lambda W_b(\lambda) }
}
where $D_L(z)$ is the luminosity distance and $g^\mathrm{AB}$ is the AB flux normalization \citep[see \eg][]{Hogg:2002yh}.

The type index implies that we use a finite set of (ordered and labeled) SEDs to describe our data. 
They will be arbitrarily normalized at a reference wavelength, here $\lambda_\mathrm{ref}=4500$ \AA.
Ideally, those templates would evolve with redshift and luminosity.
Here we adopt the simplest approach and directly redshift the SEDs and apply a multiplicative luminosity scaling. 
We will discuss those assumptions in detail below.

Our approach is to start with a base library of $\ntypes$ template SEDs, $f^\mathrm{base}_t(\lambda)$ for $t=1, \cdots, \ntypes$, and add linear corrections to them, from a library of corrections $f^\mathrm{corr}_k(\lambda)$ for $k=1, \cdots, \ncorr$. In photometric flux, since integration in photometric bands is a linear operation, we have
\eqn{
	{F}_{tb}(z) = {F}^\mathrm{base}_{tb}(z) +  \sum_{k=1}^\ncorr \alpha_{tk} {F}^\mathrm{corr}_{kb}(z), \label{modelmeans}
}
indexed for a type $t$ and a band $b$, where $\alpha_{tk}$ is the contribution of the $k$th element of the library to the $t$th type.
Note that both the base SEDs and the corrections can be arbitrarily normalized since, as explained in the next section, the multiplicative luminosity scaling will be marginalized over with a flat prior.

${F}_{tb}(z)$ is the mean model flux per type and per band. 
As apparent in \equref{eq:datalikelihood}, a (diagonal) model variance is an other ingredient of the model. 
We will also write it as a linear function of the SEDs corrections,
\eqn{
	{\omega}_{tb}(z) =  \ell \sum_{k=1}^\ncorr \alpha^\prime_{tk} {F}^\mathrm{corr}_{kb}(z).\label{modelvariances}
}
The $\ell$ multiplicative factor is essential since both the mean and the variance of the flux model need to be scaled.

In conclusion, our SED types/templates and their variances are linear mixtures of features, determined by the set of $2 \times \ncorr \times \ntypes$ coefficients $\vec{\alpha} = ( \cdots, \alpha_{tk}, \cdots, \alpha_{tk}^\prime, \cdots )$, which we will constrain to be positive for simplicity.
 In practice, we will write the model covariance matrix as $\vec{\Omega}_t(z, \vec{\alpha})$.

\subsection{2.3 The priors}

Since we have introduced parameters $z$ and $\ell$ as well as type $t$, which will need to be inferred for each object (or, in our case, marginalized over), we need to specify priors for those. 
We will assume that those priors are determined with a set of hyperparameters $\vec{\beta}$.
We factorize the full prior as
\eqn{
	p(z, t, \ell | \vec{\beta}) =  p(z|t, \ell, \vec{\beta})\ p(t | \ell, \vec{\beta}) \ p(\ell |  \vec{\beta}). \label{eq:fullprior}
}

For $p(\ell |  \vec{\beta})$, we adopt a flat, uniform prior, which will allow us to analytically marginalize over $\ell$, as shown in the next section.
One could also adopt a Gaussian mixture model and achieve analytic marginalization, although we found this lead to no improvement for the data set considered here.

$p(t | \ell, \vec{\beta})$ captures the relative abundance of types as a function of luminosity. 
It can take a physical form (\eg from a luminosity function) or be arbitrarily flexible (\eg a non-parametric mixture).
Instead of using the luminosity $\ell$, which is not observed, we will adopt a convenient proxi: the observed magnitude in one of the photometric bands $m$, taken as reference. We will ignore the noise, since it has a negligible effect with the smooth priors we adopt. Thus, we write
\eqn{
	p(t | \ell, \vec{\beta}) \ \longrightarrow\ p(t | m, \vec{\beta}) = \exp\left( a_t(m) \right)
}
$a_t(m)$ is a quadratic polynomial in the reference magnitude $m$ (of the form $x_{t1}m^2+x_{t2}m+x_{t3}$ with $x_{t1}, x_{t2}, x_{t3}$ three parameters per type $t$, given as an illustration here), and will be normalized such that $\sum_t \exp\left(a_t(m)\right) = 1  \ \forall m$.

$ p(z|t,   \ell, \vec{\beta})$ models the redshift distribution of each type. 
The main form we consider is smooth and has the main characteristics of observed galaxy distributions\footnote{Realistic galaxy redshift distributions (density of objects per unit volume) rise exponentially at low redshift due to volume effects, and smoothly decay at high redshift due to both the intrinsic cosmological abundance of objects and selection effects.} 
\eqn{
	 p(z|t, m, \vec{\beta}) =  z^{b_t(m)}\exp\Bigl(-\frac{z^{b_t(m)}}{c_t(m)}\Bigr) \times C^{-1}_t(m, \vec{\beta}) , \
}
where again $b_t(m)$ and $c_t(m)$ is a quadratic polynomial in $m$, unnormalized this time. The normalization factor to guarantee that  $\int_0^\infty p(z|t,   \ell, \vec{\beta}) \mathrm{d}z = 1$ is
\eqn{
	C_t(m, \vec{\beta}) = \frac{\bigl(c_t(m)\bigr)^{\frac{b_t(m)+1}{b_t(m)}} \Gamma\Bigl(\frac{b_t(m)+1}{b_t(m)}\Bigr) }{b_t(m)}
}
where $\Gamma(\cdot)$ is the Gamma function.

We also consider a more flexible form, a Gaussian mixture on a redshift grid of means $\bar{z}_k$ and widths $\bar{\sigma})$ for $k=1, \cdots, \mathrm{N}_\mathrm{grid}$,
\eqn{
	p(z|t, m, \vec{\beta}) = \sum_{k=1}^{\mathrm{N}_\mathrm{grid}} b_{tk}(m) \mathcal{N}( z - \bar{z}_k; \bar{\sigma}^2_k ).
}
The $b_{tk}(m)$'s are again quadratic in $m$, but normalized such that $\sum_k b_{tk}(m) = 1$. 

We adopt quadratic polynomials for the functional dependence in $m$ because they are a simple, yet non-trivial way to parametrize a smooth evolution of the priors as a function of magnitude.

$\vec{\beta}$ will just capture the parameters contained $a_t(m)$,  $b_t(m)$, and $c_t(m)$, for $t=1, \cdots,\ntypes$.

\subsection{2.4 The noise recalibration}

The final ingredient of our model is to allow for a small recalibration of the measured photometric noise, motivated by two observations: first, photometric noise is difficult to estimate for faint galaxies and its quality is known to be a function of galaxy properties (\eg morphology, color); second, it is typical for template-fitting methodologies to increase the photometric noise in order to deliver reliable photometric redshifts (we will discuss specific examples when we describe the DES SV data).  
We want to test those two assumptions and investigate whether noise estimates indeed need to be modified in order to obtain accurate photometric redshifts.
We introduce an extra flux variance added in quadrature to the measured flux variance,
\eqn {	
	\sigma^2_{ib} 	\quad &\longrightarrow& \quad \sigma^2_{ib}(\vec{\gamma}) =  \sigma^2_{ib} + \bigl( f_b (\hat{\vec{F}}_{\hspace*{-2pt}i}) \bigr)^2, \label{eq:extraflux}
	} 
where $f_b(\cdot)$ is a generic function of the measured fluxes, for each band. 
We adopt two models: 1) a quadratic polynomial of the reference magnitude, as in the previous section, 2) a neural network, taking as input all the measured photometric magnitudes, and producing the additional magnitude errors\footnote{We use a fully connected network with five hidden layers,  ten linear neurons each, performing $\textit{relu}(Ax+b)$.}.
The fact that those models use the noisy fluxes and not the true ones has a negligible effect on the result.
All extra magnitude errors are capped to $0.05$ and then converted to flux variances. 
$\vec{\gamma}$ is a vector collecting all the parameters contained in $f_b$ for $b=1, \cdots, B$.
 In practice, we will write the data covariance matrix as $\vec{\Sigma}_i(\vec{\gamma})$.

\subsection{2.5 The full posterior distribution}

We aim to infer the best hyperparameters, $\vec{\alpha}$, $\vec{\beta}$, and $\vec{\gamma}$, delivering accurate photometric redshifts for a set of noisy fluxes $\{ \hat{\vec{F}}_{\hspace*{-2pt}i} \}$.
In this work we focus on using a data set where true redshifts $\{ \hat{z}_i\}$ are also available (\ie from spectroscopy). 
However, the formalism can be straightforwardly extended to data without redshifts (which then need to be marginalized over), as discussed in our conclusions.
As stated above, we will also make use of the measured magnitude in the reference band, $\hat{m}_i$ when conditioning on $\ell_i$ in the prior (hatted quantities are observed and noisy, even though we neglect the small noise on the spectroscopic redshift and the reference magnitude).

The full posterior distribution (obtained by applying Bayes' theorem) of the model, applied to this data set, is the natural quantity to optimize for in order to obtain the best photometric redshifts possible. It reads
\eqn{
	&& p\bigl(\vec{\alpha}, \vec{\beta}, \vec{\gamma} | \{ \hat{\vec{F}}_{\hspace*{-2pt}i}, \hat{z}_i \}\bigr) \propto \label{eq:fullposterior}  \\  
	&&\hspace*{1cm} p(\vec{\alpha}, \vec{\beta}, \vec{\gamma})  \hspace*{-0.5mm} \prod_{i=1}^{\nobj}  \sum_{t_i=1}^{\ntypes} \int    p\bigl(\hat{\vec{F}}_{\hspace*{-2pt}i}(\vec{\gamma}) | \vec{F}_{t_i}(\hat{z}_i,  \vec{\alpha}), \vec{\Omega}_{t_i}(\hat{z}_i,  \vec{\alpha}), \vec{\Sigma}_i(\vec{\gamma}), \ell_i \bigr) \ p\bigl(\hat{z}_i, t_i, \ell_i | \vec{\beta}\bigr) \ \d \ell_i \quad\ \  \nonumber
}
Since we do not know the parameters of each galaxy ($t_i, \ell_i$) and they are not parameters of interest (except maybe in a post-processing stage), we have marginalized over them. 
It is clear that the hyperparameters $\vec{\alpha}$ and $\vec{\gamma}$ enter the object likelihood since they define the SED templates, while $\vec{\beta}$ enter the priors. 
We are assuming that the objects are independent, which is not formally true since there are correlations between sources found in images and across the footprint of a survey. 
We will neglect those correlations at this stage because they significantly complicate the modeling and have a negligible impact on photometric redshifts for the purposes of this work.

Let us rewrite \equref{eq:fullposterior} as
\eqn{
	p\bigl(\vec{\alpha}, \vec{\beta}, \vec{\gamma} | \{ \hat{\vec{F}}_{\hspace*{-2pt}i}, \hat{z}_i \}\bigr) \ \propto \ p(\vec{\alpha}, \vec{\beta}, \vec{\gamma}) \prod_{i=1}^{\nobj} \  \sum_{t=1}^{\ntypes} Q_{it}(\vec{\alpha}, \vec{\beta}, \vec{\gamma})  \label{eq:fullposterior2}
}
where we have introduced a \textit{marginal evidence} (or fully marginalized likelihood) per object and per type, which quantifies how well each object is modeled by each template and its priors, themselves parametrized with $\vec{\alpha}$, $\vec{\beta}$, and $\vec{\gamma}$, with the intrinsic parameters $(\ell_i, t_i)$ marginalized over. 
Those marginal evidences are calculated as
\eqn{
	Q_{it}(\vec{\alpha}, \vec{\beta}, \vec{\gamma}) &=& \int \d \ell_i  \ 
	p\bigl(\hat{\vec{F}}_{\hspace*{-2pt}i} (\vec{\gamma}) |  \vec{F}_{t}(\hat{z}_i,  \vec{\alpha}), \vec{\Omega}_t(\hat{z}_i,  \vec{\alpha}), \vec{\Sigma}_i(\vec{\gamma}), \ell_i \bigr) \ p\bigl(\hat{z}_i, t, \ell_i | \vec{\beta}\bigr) . \quad 
}
We have dropped the $i$ index in $t_i$ for conciseness.
The first term in the integral is the likelihood, and the second term the population prior.
Generally speaking, the integration over $\ell$ can be difficult to perform, and one might need to resort to sophisticated numerical integration techniques, tuned to each object.
However, for the choices we discussed above, \ie a Gaussian likelihood, a Gaussian flux model, a multiplicative $\ell$ scaling, and a flat $\ell$ prior, the integration can be analytically performed, thanks to convenient properties of Gaussians and exponentials. 
This marginalization is only approximate due to the model variance being scaled by $\ell$, but as discussed in \cite{LeistedtHogg2017}, iteratively solving for this term is a fast and accurate approximation of the true marginalization.
In addition, as discussed previously, we use the observed (noisy) magnitude of each object, $\hat{m}_i$, in the prior, \ie $p(z, t, \ell | \vec{\beta} )$, which becomes $p\left( \hat{z}_i,  t |   \hat{m}_i, \vec{\beta}  \right)$.

With those choices, we find that
\eqn{
	&&Q_{it}(\vec{\alpha}, \vec{\beta}, \vec{\gamma}) \  =\ p\left( \hat{z}_i,  t |   \hat{m}_i, \vec{\beta}  \right) \int \d \ell \ 
	\mathcal{N} \bigl(\hat{\vec{F}}_{\hspace*{-2pt}i}(\vec{\gamma})  - \ell \vec{F}_{t}(\hat{z}_i,  \vec{\alpha}) ; \vec{\Sigma}_i(\vec{\gamma}) + \ell^2\vec{\Omega}_t(\hat{z}_i,  \vec{\alpha})\bigr)    \\
	&& \approx  \  \frac{p \left( \hat{z}_i , t | \hat{m}_i, \vec{\beta}  \right)}{\sqrt{(2\pi)^B \ J_{it}(\hat{z}_i,  \vec{\alpha}, \vec{\gamma}) \ |\vec{A}_{it}(\hat{z}_i, \vec{\alpha}, \vec{\gamma})|}} \exp\left( - \frac{1}{2} G_{it}(\hat{z}_i, \vec{\alpha}, \vec{\gamma}) + \frac{1}{2} \frac{H^2_{it}(\hat{z}_i, \vec{\alpha}, \vec{\gamma}) }{J_{it}(\hat{z}_i, \vec{\alpha}, \vec{\gamma}) } \right), \quad\quad
}
where $|\cdot|$ denotes the matrix determinant. We have introduced the scalar terms
\eqn{
	 G_{it}(z, \vec{\alpha}, \vec{\gamma}) \ &=&\ \hat{\vec{F}}^T_i(\vec{\gamma}) \  \vec{A}_{it}(z, \vec{\alpha}, \vec{\gamma})^{-1}\ \hat{\vec{F}}_{\hspace*{-2pt}i}(\vec{\gamma}) \\
	 H_{it}(z,  \vec{\alpha},  \vec{\gamma}) \ &=&\  \vec{F}^T_{t}(z,  \vec{\alpha}) \ \vec{A}_{it}(z, \vec{\alpha}, \vec{\gamma})^{-1}\ \hat{\vec{F}}_{\hspace*{-2pt}i}(\vec{\gamma}) \\
	 J_{it}(z,  \vec{\alpha}, \vec{\gamma}) \ &=&\   \vec{F}^T_{t}(z,  \vec{\alpha})\ \vec{A}_{it}(z, \vec{\alpha}, \vec{\gamma})^{-1}\ \vec{F}_{t}(z,  \vec{\alpha}) 
}
as well as the total scaled photometric covariance
\eqn{
	\vec{A}_{it}(z, \vec{\alpha}, \vec{\gamma}) \ &=& \ \vec{\Sigma}_i(\vec{\gamma}) \ + \ L^2_{it}(z, \vec{\alpha}, \vec{\gamma}) \ \vec{\Omega}_t(\vec{\alpha})	
}
and the normalization term
\eqn{
	L_{it}(z, \vec{\alpha}, \vec{\gamma})  \ &=& \ \frac{\vec{F}^T_{t}(z,  \vec{\alpha}) \ \vec{\Sigma}_i(\vec{\gamma})^{-1}\ \hat{\vec{F}}_{\hspace*{-2pt}i}(\vec{\gamma}) }{  \vec{F}^T_{t}(z,  \vec{\alpha})\ \vec{\Sigma}_i(\vec{\gamma})^{-1}\ \vec{F}_{t}(z,  \vec{\alpha}) },
}
where we have kept the dependence on the hyperparameters explicit. This is a direct implementation of the approximate iterative $\ell$ marginalization (with two steps) described in \cite{LeistedtHogg2017}.
Note that the marginalization runs over $-\infty< \ell < \infty$ which is unphysical since SEDs (and thus $\ell$) are restricted to be positive. 
Also, we have extract the $\ell^2$ multiplicative term from ${\omega}_{tb}(z)$ in \equref{modelvariances}.
But for most objects, $\ell$ will either be well constrained (thus, positive), or unconstrained, in which case the evidence will be low and it will not impact any of the results.
This can be alleviated by adopting Gaussian priors as suggested above, but we found it had no effect on the results presented here.
$p\left( \hat{z}_i,  t |   \hat{m}_i, \vec{\beta}  \right)$ is simply taken from the previous section.

Note that for a galaxy where the redshift is unknown, $Q_{it}(\vec{\alpha}, \vec{\beta}, \vec{\gamma})$ is a function of redshift and the probability distribution of interest.
What we refer to as the ``redshift PDF" is $p(z) \propto \sum_{t=1}^\ntypes Q_{it}(\vec{\alpha}, \vec{\beta}, \vec{\gamma})$, normalized to integrate to one.

\subsection{2.6 Optimization}

We implemented the model presented above in \code{Tensorflow} \citep{tensorflow2015, tensorflow2016}.
We maximize $\log p\bigl(\vec{\alpha}, \vec{\beta}, \vec{\gamma} | \{ \hat{\vec{F}}_{\hspace*{-2pt}i}, \hat{z}_i  \}\bigr)$ with respect to all the parameters, captured in the generic vectors $\vec{\alpha}, \vec{\beta}, \vec{\gamma}$.
Optimization was realized with the \textit{Adam} optimizer \citep{adamoptimizer2014}, an efficient method for estimating and updating gradients and momenta to achieve stable convergence even in high numbers of dimensions.
We adopted a learning rate of $0.01$ and $50,000$ steps, which warrants convergence for all the model scenarios considered here.

Note that all of the parameters are initialized as Gaussian random numbers (zero mean and small variance, typically $0.01$) and then converted to positive numbers using exponentials or sigmoids, and also normalized properly when needed. 
As described in the next section, we run the optimization on a spectroscopic training set, and then apply the trained model on an independent validation set (also spectroscopic).

An in-depth discussion of the assumptions and limitations of this approach will be provided in our concluding section.

\section{3. Analysis of the DES SV data}\label{sec:data}

We now apply the new hierarchical model described above to the DES SV data, and compare it to existing photometric redshift estimates.

\subsection{3.1 Photometric data}

The Dark Energy Survey (DES) is a survey covering $\sim$ 5000 square degrees of the Southern sky in five optical bands, {\it grizY}.
The DES Science Verification (SV) data is a small post-commissioning survey reaching close to the depth of the full DES data, and was successfully exploited for a range of early cosmological and astrophysical studies \citep[\eg][]{CrocceDES2016, GiannantonioDES2016, DESshear2016, KacprzakDES2016, KwanDES2017}.
The associated data products were publicly released\footnote{https://opensource.ncsa.illinois.edu/confluence/display/DESDM/DES+SVA1+Data+Products}.
and consist of measurements obtained from the calibrated, co-added (or multi-epoch in some cases) images, as well as derived quantities, such as photometric redshifts, described below.

The SVA1 GOLD catalog\footnote{\url{https://des.ncsa.illinois.edu/releases/sva1}} provides \textit{griz} photometry and simple classification for objects detected in the SVA1 coadd images, covering $\sim$ 250 square degrees in about six discontinuous fields with non-uniform depth and data quality. 
The median 10$\sigma$ limiting magnitude for galaxies are approximately 24.0, 23.8, 23.0, and 22.3, in \textit{griz}, respectively. 
More detail about the catalogs and maps of the limiting magnitudes and survey properties can be found in  \cite{JarvisDES2016}, \cite{LeistedtDES2016}, and \cite{Rykoff2015}. 
In what follows we will take the $i$ band as a reference band (for all comparisons and as proxi for the absolute luminosity).

\subsection{3.2 Spectroscopic data}

To create data sets to train and validate our model on, we perform a spatial cross-match (to the nearest neighbor within $2''$) of the SV data with the following spectroscopic catalogs:
\code{VVDS}\footnote{\url{http://cesam.lam.fr/vvds/}} \citep{vvds2013}, \code{ozDES}\footnote{\url{http://www.mso.anu.edu.au/ozdes}} \citep{ozdes2017},  \code{ACES}\footnote{\url{http://localgroup.ps.uci.edu/cooper/ACES/home.html}} \citep{ACES2012}, and \code{zCOSMOS}\footnote{\url{http://cesam.lam.fr/zCosmos}} \citep{zcosmos2009}.
We do not use more spectroscopic data since for this demonstration we do not aim at having the biggest training set. 

\begin{figure}\centering
\hspace*{-3mm}\includegraphics[width=14.3cm]{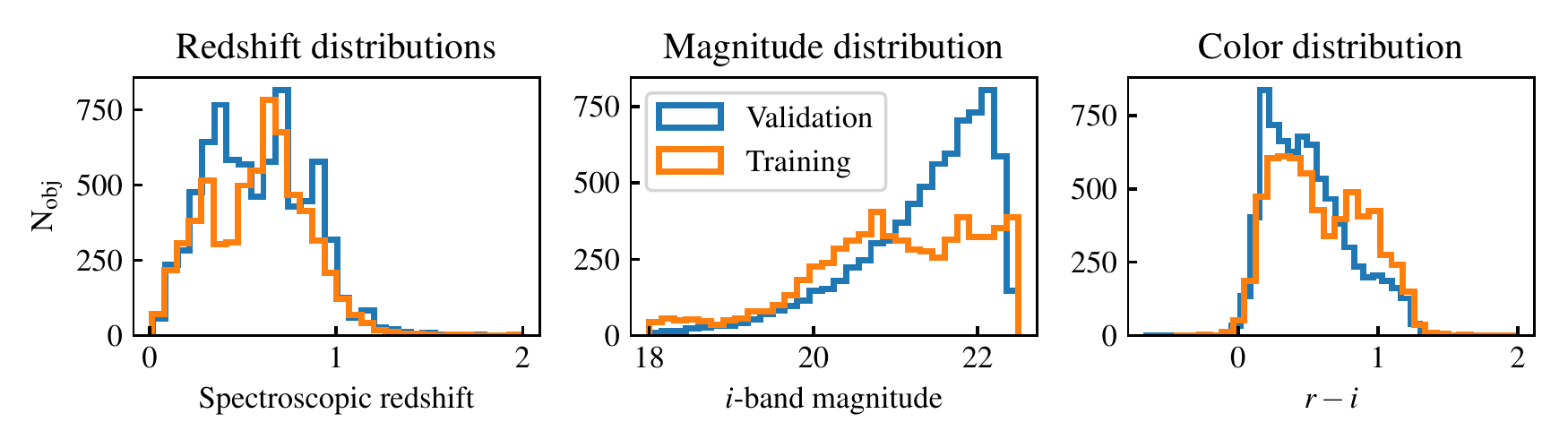}
\caption{Redshift, magnitude, and color distributions of the spectroscopic training and validation data.}\label{fig:datadistributions}
\end{figure}

We keep the 7590-object \code{zCOSMOS} cross-match for validation, and the rest (6225 objects) for training.
The redshift and $i$-band magnitudes of those training and validation sets are shown in \figref{fig:datadistributions}.
The validation set based on \code{zCOSMOS} has the fewest selection effects and cuts, and is expected to resemble the actual redshift distribution of DES SV galaxies.
The redshift and $i$-band magnitude distributions of the data are shown in \figref{fig:datadistributions}.
The training set is more heterogeneous since it is assembled from sets which have their own selection effects (note that in \citealt{Bonnett2016} a subset of the \code{VVDS} catalog was used as validation).
Even though the training and validation sets have a similar domain, they are significantly different, with the ratio of the number of objects greater than two in regions of critical interest (\eg $i>20$ and $r-i > 0.5$).

\subsection{3.3 Benchmark photometric redshifts}

Photometric redshift estimates were derived for all objects in the SVA1 GOLD catalog using a range of methods, as described in \cite{Bonnett2016}. 
We will compare the results of our hierarchical model to two of those: \bpz and \skynet. 
We choose to focus on those since they are representative of their classes (template fitting and machine learning) and were used in a range of analyses of the SV data.

\bpz \citep{Benitez:1998br} is a popular template-fitting \photoz method, that can be seen as a particular case of the model presented above, with fixed SEDs and no data corrections. 
The priors are very similar\footnote{
It uses the same flat $\ell$ prior, a similar redshift distribution $p(z  |m, t) = z^{\alpha_t}\exp\left(-(z/\beta_t(m))^{\alpha_t}\right)$ with $\beta_t(m) = \beta_t+k_{t}(m-20)$, and a similar type distribution $p(t|m) = f_t \exp\left( -g_t(m-20)\right)$. The full set of (scalar) parameters is $\alpha_t, g_t, \beta_t, k_{t}$, for $1, \cdots,\ntypes$. These are fixed for classes of SEDs (elliptical, spiral, and irregular galaxies) as well as for interpolated types.
} and are typically also calibrated on a spectroscopic data set. 
In other words, our model generalizes \bpz in the following ways: 1) the SEDs have linear corrections and a Gaussian variance, 2) the data noise is recalibrated, 3) the priors are more flexible.
Importantly, the benchmark SV \bpz results of \cite{Bonnett2016} used two interpolated templates, and set the minimal magnitude error for all objects in all four bands to be $0.05$ (\ie all smaller values were replaced by $0.05$). 
This is a special case of our noise recalibration procedure. 

\skynet is a machine-learning \photoz method using Mixture Density Networks (a neural network ending with a mixture model) to produce redshift posterior probability distributions (PDFs) directly from fluxes and flux errors. 
Those posterior distributions are, however, based on a model and a prior that are defined by the neural networks and are not directly accessible.

\begin{figure}\centering
\hspace*{-4mm}\includegraphics[width=13.cm]{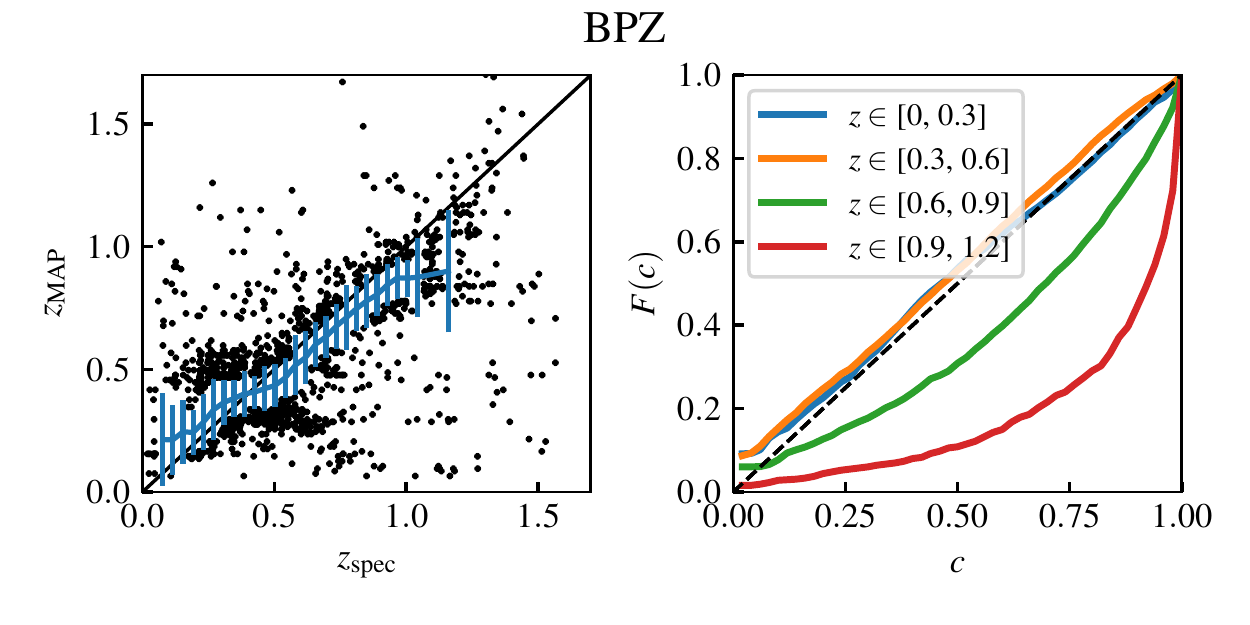}
\vspace*{-4mm}\\
\hspace*{-4mm}\includegraphics[width=13.cm]{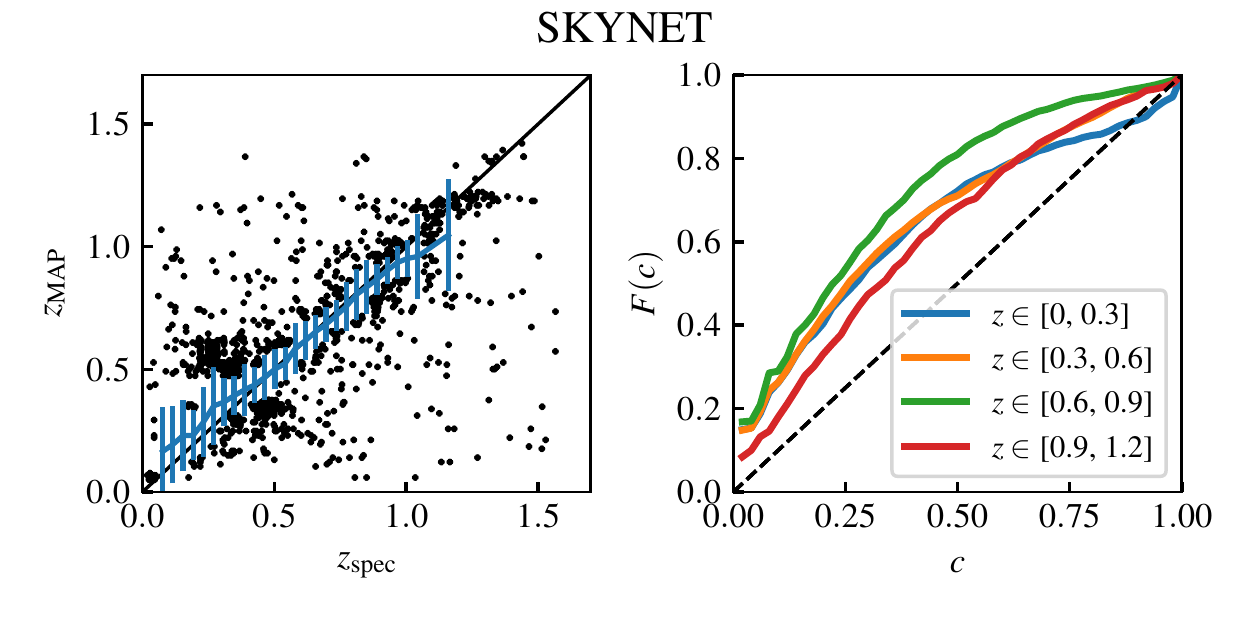}
\caption{Summary metrics for the DES SV \bpz and \skynet photometric redshifts. The left panels show the bias and scatter for the maximum a posteriori values of the redshift PDFs (\ie the peak), while the right panels show the measured $F(c)$ statistic (see text and \equref{eq:fc} for details) which should be diagonal for statistically accurate redshift PDFs. We see that \bpz yields biased \photoz's but accurate probability distributions out to $z=0.6$, while \skynet delivers unbiased \photoz's but systematically over-estimates the errors. Our hierarchical model aims at resolving both issues.}\label{fig:bpzskynet}
\end{figure}

\figref{fig:bpzskynet} shows the \bpz and \skynet results on our training and validation sets.
The left panels are scatter plots of the maximum a posteriori (\textsc{map}) value of the redshift PDF (\ie the peak) versus the true (spectroscopic) redshift. 
The line shows the mean and the errorbars the standard deviation of the sample. 
We also show the points that are outside one standard deviation.

While bias and scatter of the point estimates are useful statistics, for most cosmological applications it is the redshift distributions of galaxy samples that are of prime interest. 
Those must be derived from the redshift PDFs.
What matters for redshift distributions is whether degeneracies are correctly captured statistically in the PDFs, \ie whether the secondary features in the PDF do correspond to probabilities of finding the object at those other redshifts.
To evaluate this particular aspect of the statistical properties of the PDFs, we divide the data sets into four redshift bins (using the spectroscopic redshifts) and compute the PDF confidence intervals in each redshift bin.
For each object, we find the value of the redshift posterior distribution at the spectroscopic redshift. 
We then integrate the posterior distribution where it is greater than this value.
Specifically, for a PDF $p(z)$, we compute
\equ{
	c = \int_{\mathcal{Z}} \d z \ p(z)  \label{eq:fc}
}
with $\mathcal{Z} = \{ z \ : \ p(z)  \ge p(z=z_{\mathrm{spec}}) \}$.
For the confidence intervals arising from the redshift posterior to be statistically correct, the $c$'s should be uniformly distributed over $[0, 1]$.
We will examine the cumulative distribution $F(c)$ and compare it to the $F(c)=c$ diagonal.  
The $F(c)$ vs $c$ plot is known as a Q-Q plot.

From the four panels of \figref{fig:bpzskynet}, we can conclude that:
\begin{itemize}
\item \bpz delivers redshifts (both the \textsc{map} and the mean redshift, calculated from the PDFs) that are statistically biased low, while \skynet delivers unbiased results. This isn't unsurprising given that \bpz relies on a rigid SED model and priors, while the training and objective function in \skynet are designed to minimize bias. This was extensively discussed in \cite{Bonnett2016} for those data. 
\item Nevertheless, \bpz delivers PDFs that are more accurate statistically than \skynet: the Q-Q plots indicate that in the first two redshift bins the uncertainties are statistically accurate. They are underestimated in the last two redshift bins. For \skynet, errors are significantly overestimated in all bins. This will be evident below when we show examples of redshift PDFs.
\end{itemize}

Our hierarchical model aims at simultaneously addressing those issues: by adopting a more flexible (yet physical) model and a motivated objective function (the full posterior distribution), we hope to get more accurate results than \bpz and approach the accuracy of \skynet, while keeping the redshift PDFs fully interpretable in the context of an SED model. 

\subsection{3.4 Hierarchical model(s)}

We now describe the specifics of the hierarchical model we build and compare to \bpz and \skynet.

\begin{figure}\centering
\hspace*{-5mm}\includegraphics[width=15cm]{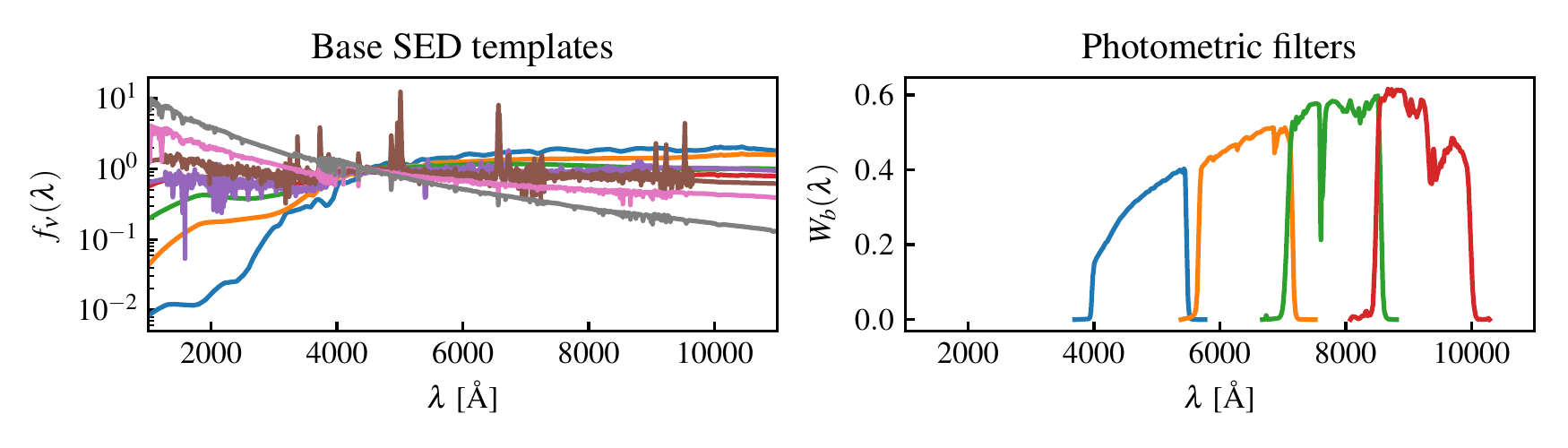}
\caption{Base template SEDs and photometric filter responses used in the hierarchical model presented here. The actual SEDs employed in the photometric redshift model include linear corrections, shown in Appendix~\ref{sec:sedscomps}.}\label{fig:sedtemplatesandfilters}
\end{figure}

\paragraph{Base SED templates} For the base library of SEDs creating our initial flux--redshift models ${F}^\mathrm{base}_{tb}(z)$, we make use the eight classic CWW templates \citep{Coleman:1980ej, Benitez:1998br} also at the center of \bpz.
We normalize the templates to $1$ at $\lambda=4,500$\AA\ and order them by their value at $2,500$\AA. The templates as well as the DECAM photometric filters are shown in \figref{fig:sedtemplatesandfilters}.
In addition, we introduce extra templates resulting from linear interpolation between successive base templates.
We will explore the effect of varying the number of interpolated templates.

\paragraph{SED corrections and variance} For our linear corrections, ${F}^\mathrm{corr}_{kb}(z)$, we would like to incorporate sensible features that are found in real or model spectra. To achieve this, we perform a non-negative matrix factorization (PCA with positive components or eigenvectors) of a large library of templates from the \code{PEGASE} stellar population synthesis modeling framework \citep{pegase1, pegase2}. We use the grid shipped with the \textsc{eazy} \citep{EAZY2008} code\footnote{https://github.com/gbrammer/eazy-photoz/blob/master/templates/pegase13.spectra.param}, initially described in \cite{Grazian2006}. 
We run the non-negative matrix factorization with 76 components, shown in the first subplots of \figref{fig:sedcorrections}.
The remaining subplots are extra components we manually add: 24 Gaussians on a logarithmic grid between $1,000$\AA\ and $10,000$\AA.
As discussed previously, these features will be used to linearly correct the base templates (including the interpolated ones) and add variance to the model SEDs and flux--redshift models.

\subsection{3.5 Results}

Table \ref{table:results} shows the grid of models we optimized on the training data\footnote{As mentioned above, we use a learning rate of $0.01$ and $50,000$ iterations with the \textit{Adam} optimizer. Unsurprisingly, execution time mostly depends on the number of objects and the number of parameters. For our training data, the smallest models take about $0.01$ second per iteration, while the more complex ones go up to a few seconds per iteration. This is not a limiting factor here, but this could be addressed in various way, for example by using stochastic optimization. Computing the redshift PDFs at fine resolution takes about twenty times longer.}.
We have a column for $\mathrm{N}_\mathrm{par}$, the number of parameters of each model.
The last two columns show the total values of the posterior distribution of \equref{eq:fullposterior} on the training and validation data, divided by the number of objects, to facilitate comparison.
Unsurprisingly, it is systematically higher for the validation data (since the training procedure does not know about those objects, and they are fairly different to the training set).
In what follows we draw a number of conclusions from these runs. Although we have compared all models in detail, we only include a few relevant figures and comparisons in this paper (the rest are available upon request).

\begin{table}\centering
\footnotesize
\begin{tabular}{|c|c|c|c|c|c|c|c|c|}
\hline		&	interp 	&	prior	&	SED mean 	&	SED 	
	&	mag error  & $\mathrm{N}_\mathrm{par}$ & $\log [Q]/\mathrm{N}_\mathrm{obj}$ & $\log [Q]/\mathrm{N}_\mathrm{obj}$ 	\\
&	 SEDs	&	$p(z, t, m)$ 	&corrections	&	 variances	
 &	 	corrections	& &	 (training)	&	 (validation)  	\\\hline
A & 2 & simple & \checkmark &  &  & 2398 & -10.18 & -5.66 \\ 
B & 2 & simple &  &  &  f(m) & 210 & 17.53 & 18.34 \\ 
C & 2 & simple & \checkmark &  &  f(m) & 2410 & 19.49 & 20.07 \\ 
D & 0 & simple & \checkmark & \checkmark &  & 1672 & 18.64 & 19.32 \\ 
E & 2 & simple & \checkmark & \checkmark &  & 4598 & 19.86 & 20.42 \\ 
F & 2 & GMM & \checkmark & \checkmark &  & 5126 & 19.84 & 20.27 \\ 
G* & 2 & simple & \checkmark & \checkmark &  f(m) & 4610 & 19.91 & 20.46 \\ 
H & 2 & GMM & \checkmark & \checkmark &  f(m) & 5138 & 19.85 & 20.31 \\ 
I & 2 & simple & \checkmark & \checkmark &  NN & 5022 & 19.82 & 20.36 \\ 
\hline	
\end{tabular}
\caption{Summary of the models we investigated and optimized the parameters of. For the extra photometric noise, $f(m)$ refers to the quadratic polynomials in the reference magnitude, while NN refers to the neural network corrections, functions of the magnitudes in all bands. Model G is marked with a star because it will be the focus of most of the discussion and figures below.}\label{table:results}
\end{table}

\begin{figure}\centering
\vspace*{-1mm}\hspace*{-4mm}\includegraphics[width=13cm, trim = 0cm 0.cm 0cm 0.6cm, clip]{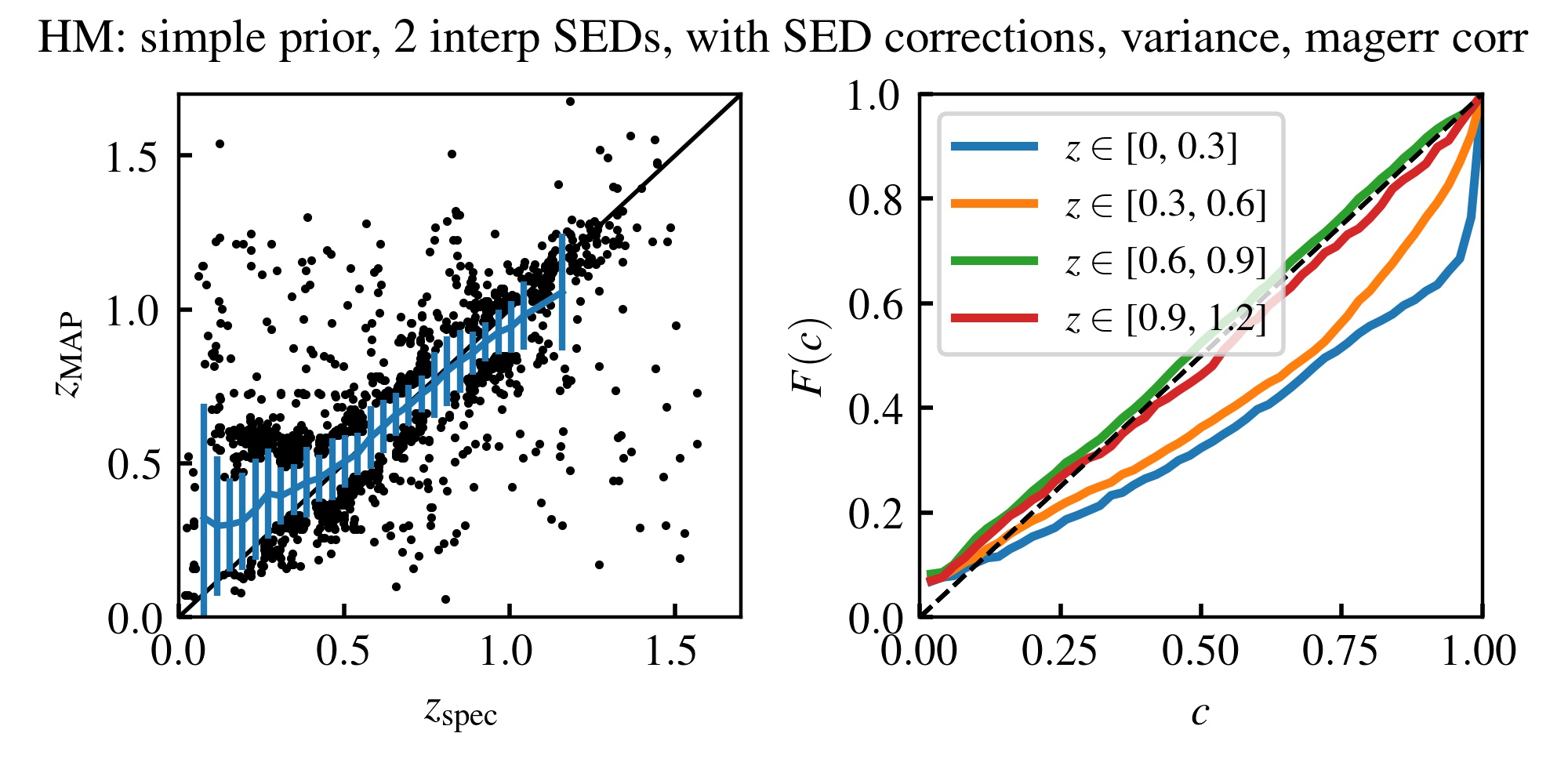}
\caption{Same as \figref{fig:bpzskynet} but for one of the hierarchical models of Table \ref{table:results}, model G, which includes SED corrections and variance, and magnitude error corrections. Not only does it provide unbiased point estimates with small scatter for $z > 0.6$, but the redshift PDFs in this range are also statistically accurate as measured by diagonal $F(c)$ estimates. Inaccuracies at low redshift are due to an incomplete SED basis.}\label{fig:zmap_vs_zspec}
\end{figure}

Our first main conclusion from Table \ref{table:results} and \figref{fig:zmap_vs_zspec} is that without introducing magnitude errors or SED variance (model A), the results are poor: errors are significantly underestimated, outlier fractions are high, and the total posterior probabilities are low. 
Those models are unable to produce reliable photometric redshifts.
This isn't surprising and partly explain why the publicly released \bpz analysis set the minimum magnitude errors to $0.05$, in order to alleviate this effect.
Note that increasing the number of interpolated templates does not improve the results.

Our second major conclusion is that introducing any kind of extra variance (SED or photometric noise) elevate the results to an acceptable accuracy (models D-I).
In fact, all models with SED and/or photometric variance roughly yield similar performance (measured from the total posterior probability, the redshift PDFs, the outlier fractions, etc).
For this reason, for most of the following figures and discussion we will focus on one of the models, model G, which has two interpolated templates, SED corrections and variance, and magnitude error corrections. 
Other models with variance give similar performance and also similar redshift PDFs.

\figref{fig:zmap_vs_zspec} shows that the Q-Q plots of those models are satisfying and almost diagonal. 
It is noticeable that the high redshift bins are more diagonal and accurate, in the regime where most of the cosmological information is, which is relevant for \eg clustering and cosmic shear studies.
The outliers are also very similar to those in the \bpz and \skynet results.
In fact, most of those objects are outliers in common between all methods and models. 
Those might be true statistical outliers, in the sense of atypical objects, for instance galaxies having lived an atypical formation and evolution. 
Some other objects may have corrupted flux or flux error measurements, for example due to blending of multiple objects, overlay with stellar streaks or cosmic rays or various other possible image features that would result in a failure or biasing of the measurement.
Investigating those outliers is beyond the scope of this paper and an active research area.

\begin{figure}\centering
\hspace*{-2mm}\includegraphics[width=15cm, trim = 0cm 0.cm 0cm 1cm, clip]{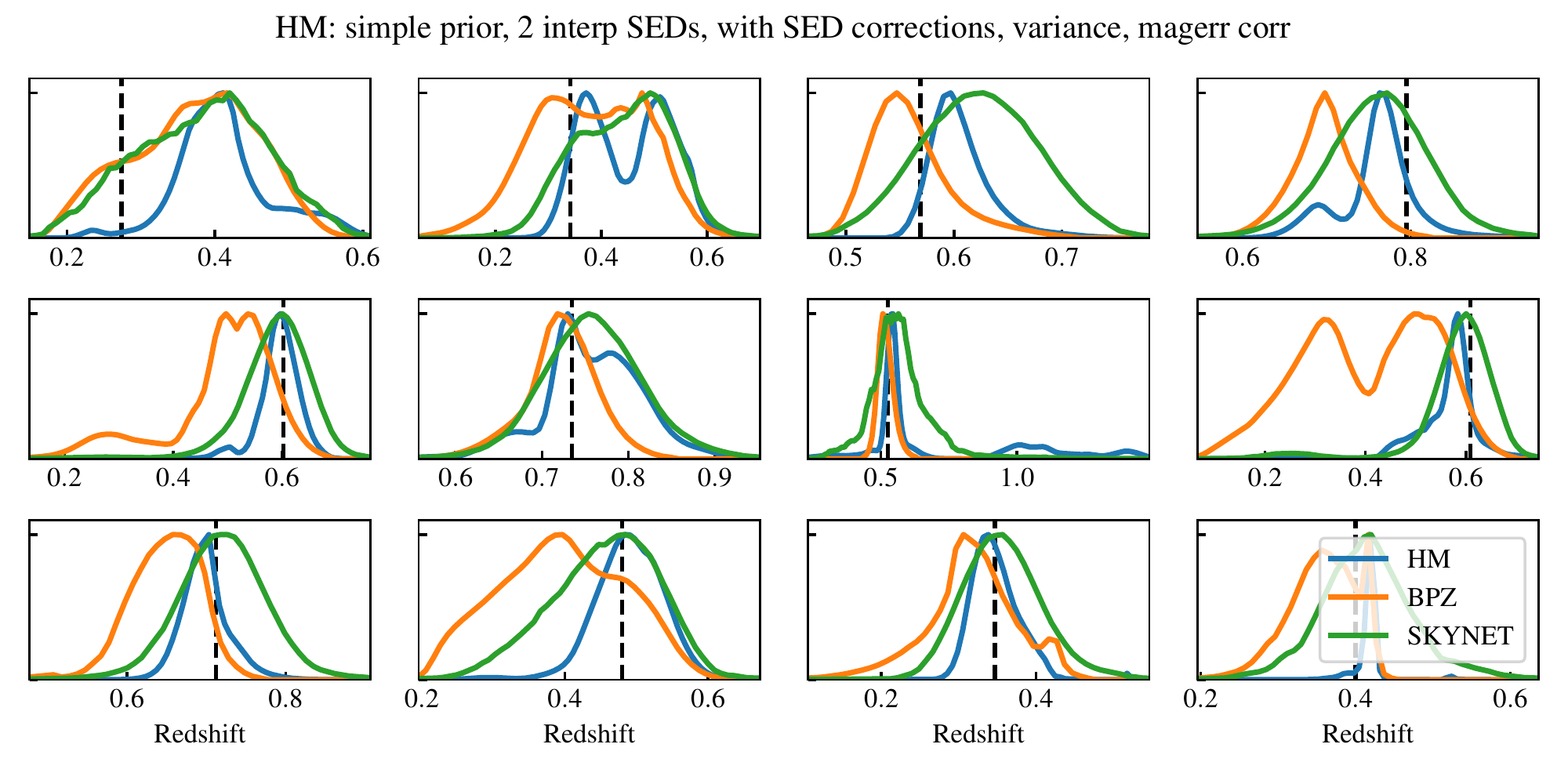}
\caption{Redshift probability distributions (obtained with model G) for twelve randomly chosen objects of our validation sample. The PDFs from the hierarchical model shares features with both \bpz and \skynet, resulting from its template-fitting nature and also its flexibility. They are also more compact (\ie more precise), as quantified in \figref{fig:zmapstds}.}\label{fig:pdfs}
\end{figure}

\begin{figure}\centering
\hspace*{8mm}\includegraphics[width=15.5cm, trim = 0cm 0.cm 0cm 0.7cm, clip]{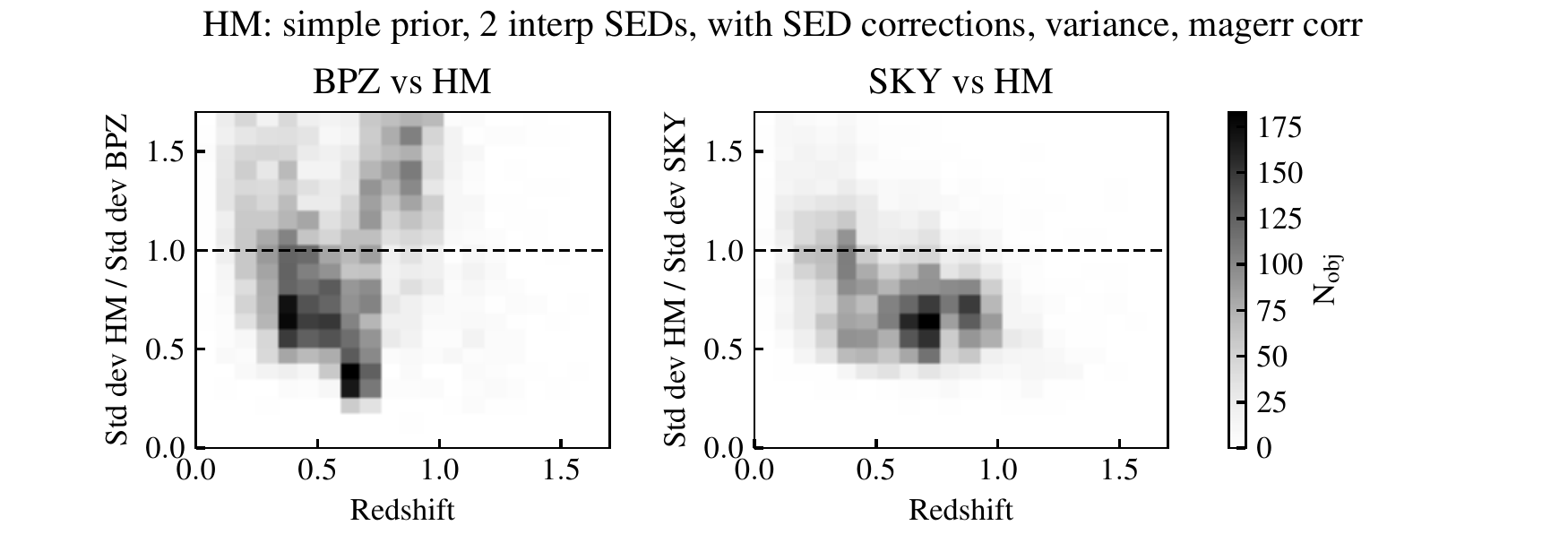}
\caption{Histograms of the ratio between the standard deviation (\ie the support) of the \skynet, \bpz, and hierarchical model PDFs, showing that the latter are significantly narrower for most objects (here for the validation set).}\label{fig:zmapstds}
\end{figure}

Redshift PDFs for the considered model are shown in \figref{fig:pdfs}.
There are striking similarities between the hierarchical model PDFs and those of \bpz and \skynet.
One the one hand, the hierarchical model is by nature a template-fitting model and its base templates are the same as \bpz. 
It is not surprising to find similar features between the two, especially in terms of secondary modes in the PDFs.
On the other hand, the hierarchical model requires a smaller amount of extra variance, and some SED variance. 
Most PDFs are much narrower than those of \bpz. This is quantified in \figref{fig:zmapstds} where we show histograms of the ratio between the standard deviation of the PDFs as a function of redshift. 
Even though this is not the most reliable point of comparison because of the multimodality of most PDFs, we can still conclude that the hierarchical model delivers PDFs that are much more compact than \bpz.
Upon examination, we find that many of them closely approach the \skynet PDFs. 
This is interesting because \skynet, as a flexible machine-learning model, attempts to exhaust the amount of information in the data and delivers maximally compact PDFs. 
The fact that the hierarchical model is able to reach similar performance (also seen in \figref{fig:zmapstds}) indicates that it is also almost exhausting the information in the data.

\figref{fig:minmagerrs} shows the extra magnitude errors added in quadrature to the measured photometric noise for three models (C, G, and I). The first two rows are for models with quadratic functions of the reference $i$-band magnitude, while the third panel is for a neural network taking all noisy magnitudes as input and computing the extra magnitude errors in all bands. 
For the latter we only show one-dimensional projections: the mean and standard deviation of the extra magnitude noise as a function of the reference magnitude. 
Two-dimensional projections are shown in Appendix~\ref{sec:appnoise}.
We see that for all three models the $g$ band systematically requires extra noise. 
Extra noise is required in the other bands, except when SED variance is part of the model, in which case only some $z$-band variance is needed.
This is also true when moving to multidimensional corrections provided by the Neural Network. 
Interestingly, the noise in the $z$ band is structured, and only needed in some regions of color space, as shown in Appendix~\ref{sec:appnoise}.
The lack of structure in the other bands and also the minor improvement in the overall posterior distribution shows that despite the more complex $z$-band noise those more flexible corrections are not necessary for those data.

\begin{figure}\centering
\hspace*{-2mm}\includegraphics[width=13.cm]{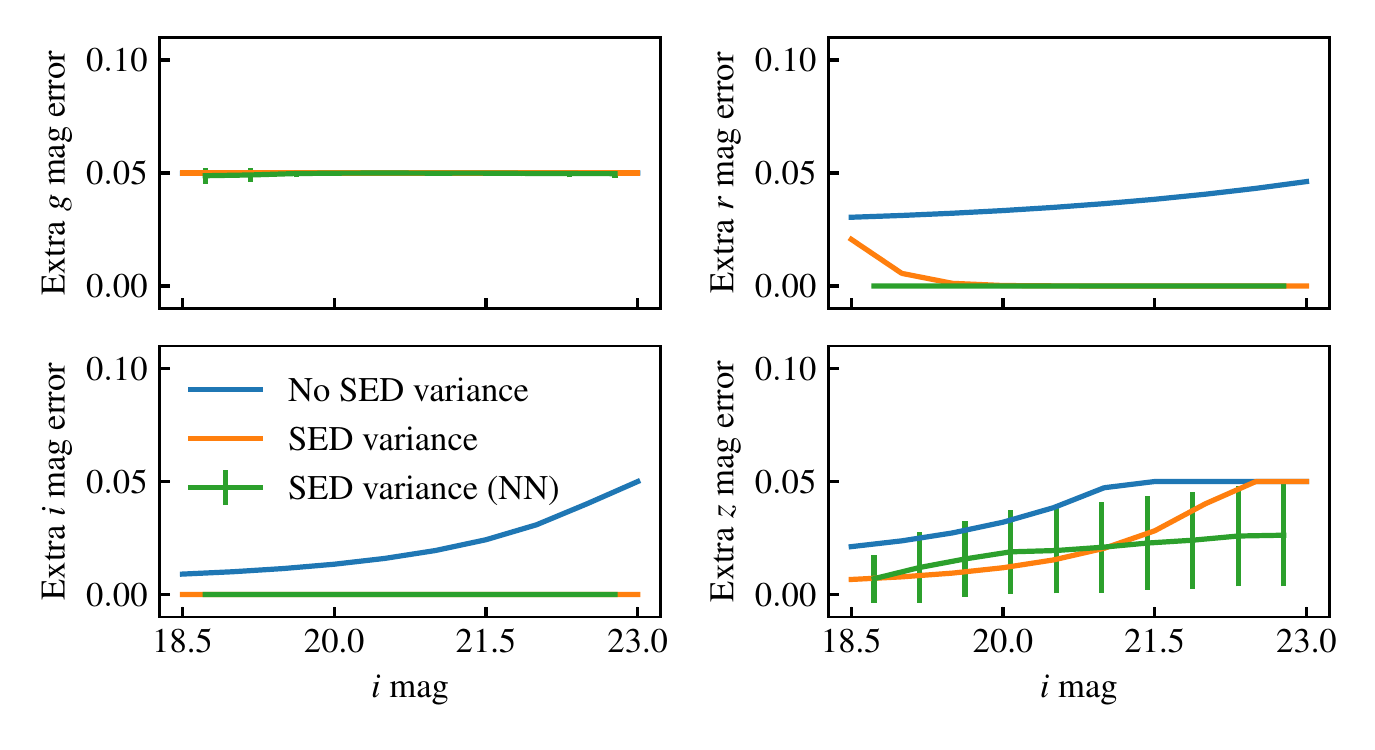}\label{fig:minmagerrs}
\caption{Extra magnitude errors added in quadrature to the measured photometric noise as described in \equref{eq:extraflux}, for three models of interest (C, G and I), all with SED corrections. 
Introducing SED variance alleviates the need for extra noise in all bands. 
This is also true for the model with Neural Network noise corrections, even though the extra flexibility of the model does not bring a huge improvement, as demonstrated in Table \ref{table:results}.
In this case, even though the full model operates in four-dimensional flux space, we only show a projection of the mean and standard deviation as a function of the $i$ band for comparison with the other two models.  
Two-dimensional projections are shown in Appendix~\ref{sec:appnoise} and have some interesting structure. 
}\label{fig:minmagerrs}
\end{figure}

\begin{figure}\centering
\hspace*{-2mm}\includegraphics[width=15.5cm]{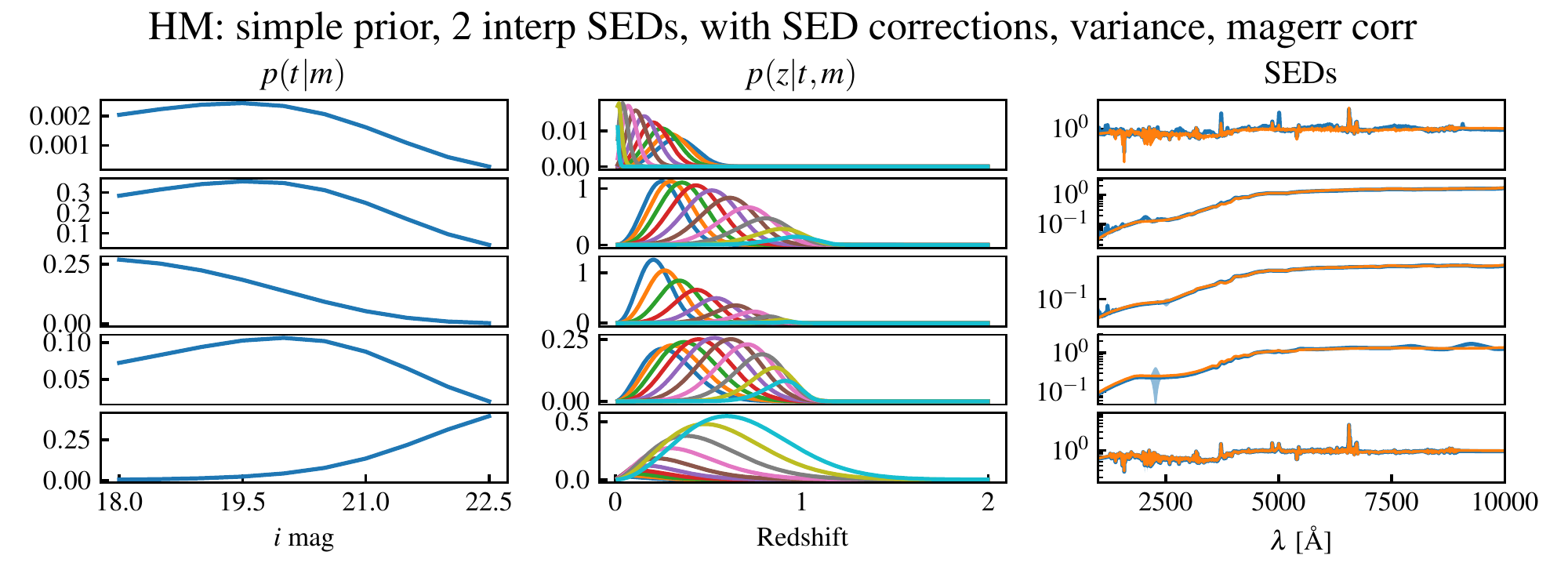}
\hspace*{-2mm}\includegraphics[width=15.5cm]{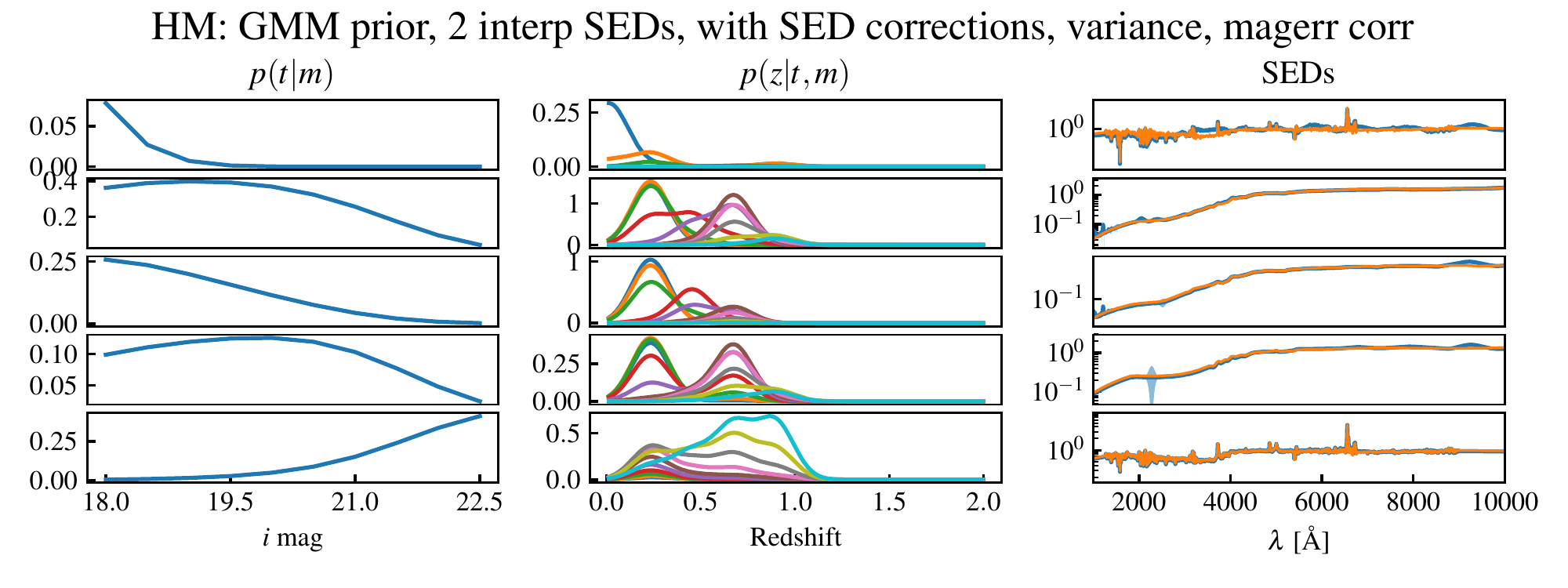}
\caption{SEDs and priors obtained for two of our models (G and H), as described in \equref{modelmeans} and \equref{eq:fullprior}, respectively. The ten colored lines in $p(z|t, m)$ correspond to a uniform tiling of the magnitude $m$ in the range $[18, 22.5]$. In the right panels, the original SEDs are in orange, and the corrected ones are in blue. Despite the corrections being small they are needed to obtain unbiased redshift estimates, as demonstrated in Table \ref{table:results}. We see that the second model, despite having many more degrees of freedom and parameters thanks to a Gaussian Mixture redshift prior, delivers very similar SEDs and priors (as well as other metrics), indicating that this flexibility is not needed to deliver accurate redshift PDFs.}\label{fig:sedsandpriors}
\end{figure}

Finally, \figref{fig:sedsandpriors} shows the SEDs and priors obtained for two hierarchical models, both with SED corrections and variance, and magnitude noise (models G and H).
The first is based on our simplest priors, while the second has the flexible Gaussian Mixture prior.
We have selected the first most significant types, as measured by the maximum value of $p(t|m)$.
We show the type and redshift components of the prior, as well as the SEDs, both the initial and corrected ones, and their variance.
First, we see that both the SEDs and the priors are very similar between the two models, despite the second having many more degrees of freedom and parameters.
This indicates that the optimization procedure is robust, and the similarity between the simple and the Gaussian Mixture priors hints that the flexibility of the latter is not really needed.
Indeed, we saw previously that those two models achieve similar performance, despite the much greater flexibility of the second.
Finally, we see that the SEDs are not corrected in any significant manner; the corrections and the variance are barely visible. 
Yet, they make a significant different in the performance of the model and the resulting redshifts, partly due to the high precision (small errors) of the photometry we use.

\section{4. Discussion and conclusion}\label{sec:concl}

We presented a hierarchical modeling approach to photometric redshifts, generalizing template fitting using flexible models for the SEDs, their priors, as well as photometric noise corrections.
We applied it to the DES SV data and compared the results of two standard algorithms, \bpz and \skynet.
Interestingly, by successfully exploring the activation or deactivation of components in the hierarchical model, we are able to gain intuition about the existing results as well as about the data themselves. 
We found that SED corrections and SED variance and/or extra photometric noise are necessary to yield accurate, unbiased \photoz's, especially at high redshift, which \bpz does not deliver. 
Models with variance perform roughly equally well and almost match the precision of \skynet, while also delivering accurate redshift PDFs (\skynet delivers over-estimated redshift errors). 
Extra model flexibility such as Gaussian Mixture priors is not required to improve the results.
Without SED variance and/or extra photometric noise, it is not possible to derive acceptable \photoz's, at least with the library of templates and the data we considered.
Interestingly, introducing SED variance alone almost entirely removes the need to increase the measured magnitude errors.
When magnitude errors are added, it is sufficient for them to be smooth functions of the reference magnitude, although we find that they are only truly needed in compact portions of colors space in the $g$ band. 

We now discuss the main limitations of our approach in detail: 

\paragraph{Model and data} 
The hierarchical model relies on a small and fixed set of base SED templates, which do not evolve with redshift, and linear corrections as well as a Gaussian variance, both constructed from a (fixed) library of features.  
Our main priors are smoothly changing as a function of redshift and luminosity.
All of these assumptions limit the set of real galaxies we can represent as a function of redshift. 
Outliers are either due to this limitation or to corrupted photometric data, which we did not attempt to model or address here.
In future work we will explore the use of richer spaces of features for the SEDs, possibly connected to stellar population models, as well as physically motivated priors connected to luminosity functions with observational selection effects.
We will also explore more sophisticated models to correct biases in photometric measurements (possibly as a function of position and survey observing conditions) and deal with outliers due to corrupted data.
We note that we have performed this study in a deep field (\ie with lower noise level than the rest of the survey), where \photoz algorithms typically perform less well due to the increase modeling accuracy needed to explain the data. 
Yet, the conclusions of this study and the comparison to results from other algorithms in the same field are not affected by this.
For the same reason, we limited ourselves to relatively bright objects ($i$-magnitude limit of 22.5) for which \bpz and \skynet were also optimized.

\paragraph{Training and optimization} 
We trained and validated the hierarchical models on spectroscopic data sets for computational efficiency and simplicity.
Given that their redshift distributions were comparable to typical target photometric galaxy catalogues, it is unlikely that this strongly biases the priors or SEDs, even though it is formally a possibility.
It is possible (even though more computationally intensive) to run the method on data without spectroscopic redshifts, where the true redshifts are then marginalized over. 
This is left for future work; as is the inclusion of extra photometric bands when available, for example in fields covered by other surveys.
Finally, we optimized the hyperparameters of the model, but one could in principle span (\eg sample) the entire space of parameters compatible with the data, and marginalize over them, although we suspect this will have a minor effect on the results.

\paragraph{Metrics and results}
We looked at a restricted number of metrics, focusing on scatter and Q-Q plots to evaluate the quality of the redshift PDFs.
In particular, we did not look at redshift distributions, even though those are a critical element needed for cosmological analyses.
This is because turning a set of redshift PDFs into redshift distribution estimates and uncertainties is not trivial and requires additional infrastructure not directly related to the process of deriving individual \photoz's \cite[see \eg][]{LMP2016} . 
In addition, this process is only meaningful if performed in a realistic survey setup, such as for a million-object catalog with spatially-varying observing conditions, as well as a validation procedure for the redshift distribution, which is not trivial given the small amount of spectroscopy available.
For those reasons, the derivation of accurate redshift distributions from the \photoz's obtained with hierarchical models is left for future work.

In summary, in this paper we have shown how some of the limitations of existing photo-$z$ methodologies can be elegantly addressed with hierarchical modeling, and how this approach opens new perspectives for meeting the stringent \photoz requirements of ongoing and future surveys. 

\acknowledgments

We thank Dustin Lang and Martin P. Rey for very useful comments on this manuscript.

BL was supported by NASA through the Einstein Postdoctoral Fellowship (award number PF6-170154).
DWH was partially supported by the NSF (AST-1517237) and the Moore--Sloan Data Science Environment at NYU.
RHW and JD received partial support from NASA ROSES ATP 16-ATP16-0084 and from the U.S. Department of Energy under contract number DE-AC02-76SF00515.

\bibliography{bib}

\appendix

\section{Components of the linear corrections and variance}\label{sec:sedscomps}

\begin{figure}\centering
\hspace*{-5mm}\includegraphics[width=15.5cm]{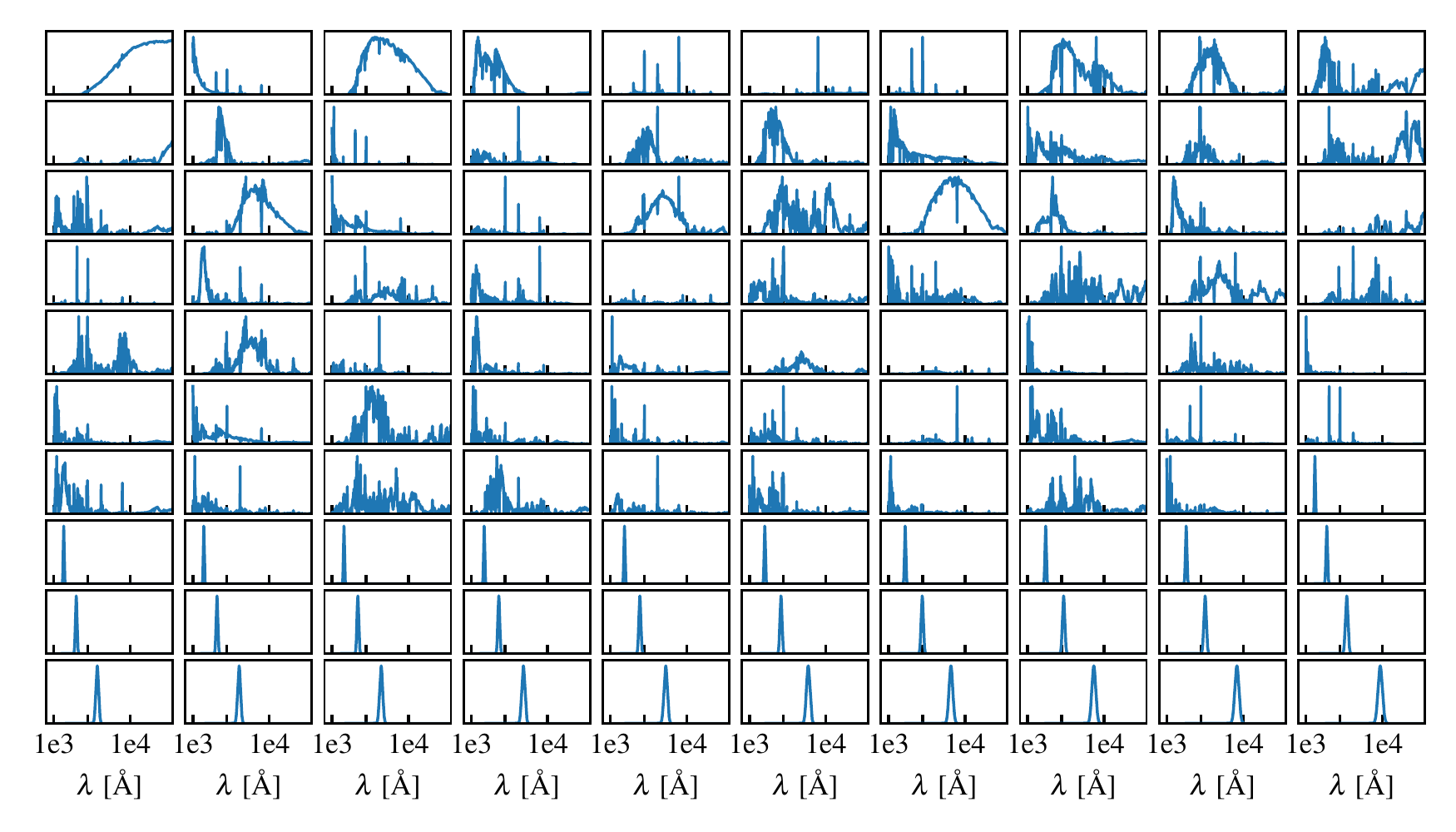}
\caption{Features to be linearly combined into SED corrections and variance. Those are the result of a non-negative matrix factorization of a large library of synthetic SEDs, complemented with a few narrow Gaussian components.}\label{fig:sedcorrections}
\end{figure}

The components of the linear corrections and variance terms, going in \equref{modelmeans} and \equref{modelvariances} are shown in \figref{fig:sedcorrections}.

\section{Flexible noise corrections with neural networks}\label{sec:appnoise}

\begin{figure}\centering
\vspace*{3mm}\hspace*{-2mm}\includegraphics[width=15.5cm, trim = 1.8cm 2.5cm 2cm 0.2cm, clip]{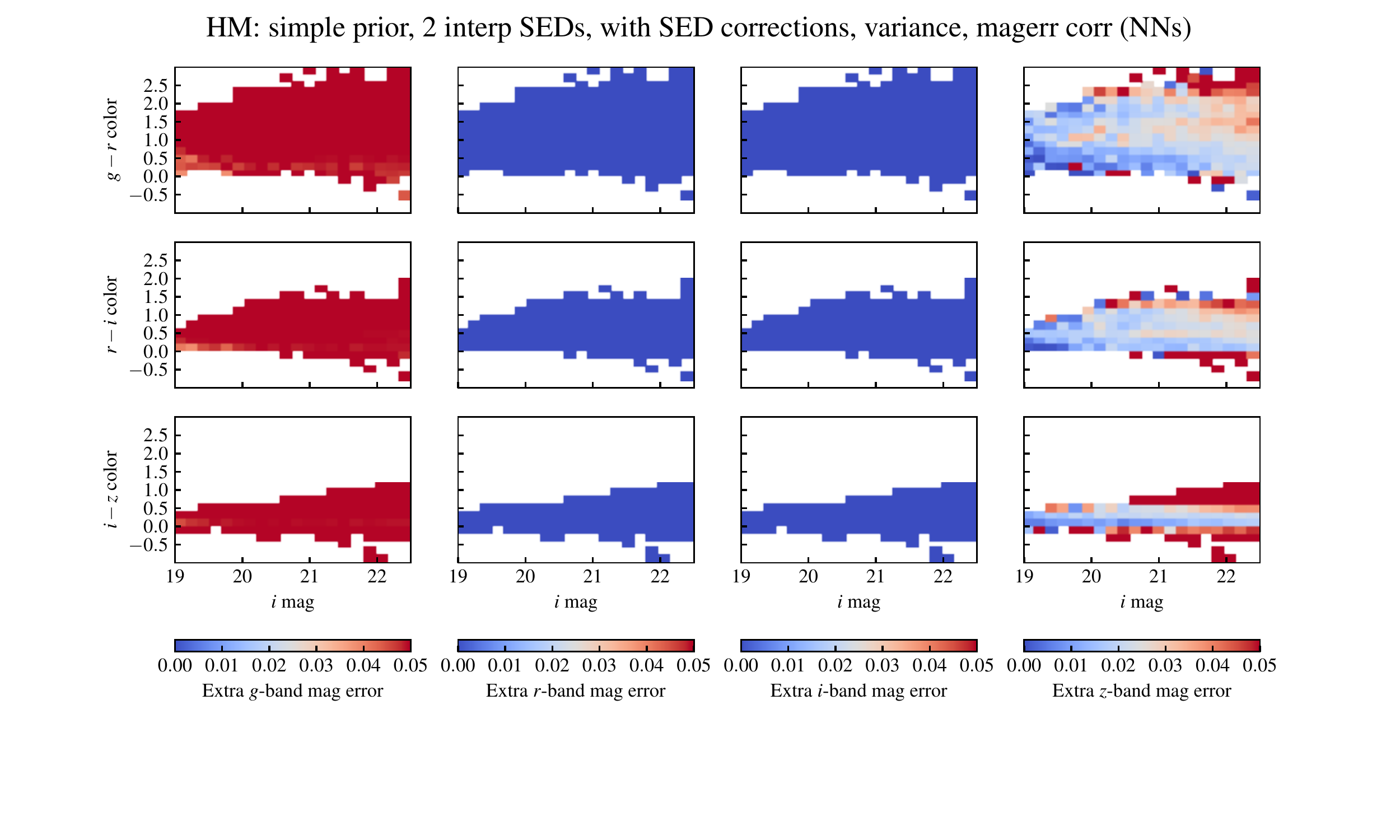}\\
\caption{Extra magnitude errors added in quadrature to the measured photometric noise. 
We feed both the training and validation data to a neural network, and compute the mean resulting extra magnitude noise in two-dimensional color-magnitude space for simplicity. 
Compared to the simpler models, the network has the same behavior in the $gri$ bands, but shows some structure in the $z$ band, where the noise is only needed in some regions of colors space. 
}\label{fig:minmagerrs_nn}
\end{figure}

One of our models has noise corrections (added in quadrature to each band) that are the output of the neural network, taking the four fluxes as inputs.
Since we cannot visualize this four-dimensional space, \figref{fig:minmagerrs_nn} shows two dimensional projections of the mean magnitude noise produced by the network, as a function of colors. 
Compared to the simpler models, the network has the same behavior in the $gri$ bands, but shows some structure in the $z$ band, where the noise is only needed in some regions of colors space. 
Nevertheless, those more complex corrections lead to a minor improvement in the posterior distribution.

\end{document}